\documentclass[ped,preprint,linenumbers,amssymb,amsmath,aps,showpacs,nofootinbib,superscriptaddress]{revtex4}
\usepackage{graphicx}
\usepackage{bm}
\usepackage{amssymb}
\usepackage{subfigure}
\usepackage{color,soul}
\setulcolor{red}

\newcommand{\beq}{\begin{equation}}
\newcommand{\eeq}{\end{equation}}
\newcommand{\beqa}{\begin{eqnarray}}
\newcommand{\eeqa}{\end{eqnarray}}

\begin{document}

\title{New Bounds on Axion-Like Particles From the Fermi Large Area Telescope observation of PKS $2155-304$}

\author{Cun Zhang}
\affiliation{School of Physics, Nanjing University, Nanjing, 210092, China}
\affiliation{Key Laboratory of Dark Matter and Space Astronomy, Purple Mountain Observatory, Chinese Academy of Sciences, Nanjing 210008, China}
\affiliation{Joint Center for Particle, Nuclear Physics and Cosmology, Nanjing 210093, China}
\author{Yun-Feng Liang}
\affiliation{Key Laboratory of Dark Matter and Space Astronomy, Purple Mountain Observatory, Chinese Academy of Sciences, Nanjing 210008, China}
\author{Shang Li}
\affiliation{Key Laboratory of Dark Matter and Space Astronomy, Purple Mountain Observatory, Chinese Academy of Sciences, Nanjing 210008, China}
\author{Neng-Hui Liao}
\affiliation{Key Laboratory of Dark Matter and Space Astronomy, Purple Mountain Observatory, Chinese Academy of Sciences, Nanjing 210008, China}
\author{Lei Feng}
\email{Corresponding author: fenglei@pmo.ac.cn}
\affiliation{Key Laboratory of Dark Matter and Space Astronomy, Purple Mountain Observatory, Chinese Academy of Sciences, Nanjing 210008, China}
\author{Qiang Yuan}
\email{Corresponding author: yuanq@pmo.ac.cn}
\affiliation{Key Laboratory of Dark Matter and Space Astronomy, Purple Mountain Observatory, Chinese Academy of Sciences, Nanjing 210008, China}
\author{Yi-Zhong Fan}
\email{Corresponding author: yzfan@pmo.ac.cn}
\affiliation{Key Laboratory of Dark Matter and Space Astronomy, Purple Mountain Observatory, Chinese Academy of Sciences, Nanjing 210008, China}
\author{Zhong-Zhou Ren}
\email{Corresponding author: zren@nju.edu.cn}
\affiliation{School of Physics, Nanjing University, Nanjing, 210092, China}
\affiliation{Joint Center for Particle, Nuclear Physics and Cosmology, Nanjing 210093, China}

\begin{abstract}
The axion-like particle (ALP)-photon mixing in the magnetic field around
$\gamma$-ray sources or along the line-of-sight could induce oscillation
between photons and ALPs, which then causes irregularities in the
$\gamma$-ray spectra. In this work we try to search for such spectral
irregularities in the spectrum of PKS $2155-304$ using 8.6 years of the
Fermi Large Area Telescope (Fermi-LAT) data. No significant evidence for
the presence of ALP-photon oscillation is obtained, and the parameter
space of ALPs is constrained. The exclusion region sensitively depends
on the poorly known magnetic field of host galaxy cluster of PKS $2155-304$.
If the magnetic field is as high as $\sim 10~{\rm \mu G}$, the ``hole"-like
parameter region allowed in Ref.~\cite{Fermi} can be ruled out.

\end{abstract}
\maketitle

\section{Introduction}

As a generalization of the axion~\cite{axion1,axion2,axion3,axion4,axion5},
the axion-like particle (ALP) is a very light pseudo-Goldstone boson
predicted in several extensions of the Standard Model, such as the
string theory~\cite{alp-string1,alp-string2} and the Kaluza-Klein
theory~\cite{alp-kk,alp-kk1,alp-kk2}. Thanks to its interaction with
the electromagnetic field, an ALP could oscillate with a photon in an
external electric or magnetic field, providing a promising way to detect it.
Different from the axion, there is no relation between the ALP mass and
its coupling strength which was employed to solve the Strong CP
Problem~\cite{axion1}.

ALPs belong to one kind of cold dark matter candidates known as weakly
interacting slim particles (WISPs)~\cite{ALP-dm,WISPs2,WISPs3}.
Other than the search for the classical weakly interacting massive particles
(WIMPs) dark matter~\cite{WIMPs1,WIMPs2,WIMPs3,WIMPs4,WIMPs5}, the detection
of ALPs is usually based on the ALP-photon oscillation effect~\cite{CAST,CAST1,ALPSI}.
A high or large-scale electric or magnetic field is needed in such a kind of experiments.
Several ground experiments, such as the Any Light Particle Search (ALPS)
I~\cite{ALPSI} and the CERN Resonant WISP Search (CROWS)~\cite{CROWS}, have
searched for such oscillation signals as the photon-ALP beam flying through
the laboratory magnetic field.
The CERN Axion Solar Telescope (CAST) and the proposed International
Axion Observatory (IAXO) instead aim to detect the photons converted
by solar axions (produced by the Primakeoff effect as photons pass the
Coulomb field of charged particles in the Sun) in the laboratory magnetic
field. The relevant experimental progresses can be found in some recent review
articles~\cite{ALPex0,axion-det1,axion-det2,axion-det3}.

Magnetic field exists everywhere in the Universe. For some highly magnetized neutron stars (i.e.,
magnetars,~\cite{Magnetars}), the dipole
magnetic fields could be high up to $\geq 10^{14}$ G. There are also large-scale
coherent magnetic fields in galaxies and galaxy clusters, with coherent
lengths as long as tens of kpc. These high magnetic field objects or
large-scale coherent regions are ideal targets to search for ALPs.
Chelouche {\em et~al.}~\cite{photon-part} calculated the spectral
signatures induced by ALP-photon oscillations in several astrophysical
scenarios. The H.E.S.S collaboration used the data of PKS $2155-304$ to
constrain the ALP parameter space~\cite{hess-alp}. The Fermi-LAT
collaboration studied possible spectral irregularities in the $\gamma$-ray
data of NGC 1275, and got strong constraints~\cite{Fermi}. However,
there is a ``hole"-like region survived in their constraints~\cite{Fermi}.
In this work, we search for spectral irregularities using the Fermi-LAT data
of PKS $2155-304$. The uncertainties of the magnetic fields, including
both the distributions and the strengths, are discussed. With ``optimistic"
parameters, the ``hole"-like region can be ruled out.

This work is organized as follows. In Section \ref{sec:model}, we briefly
describe the ALP-photon oscillation model.  In Section \ref{sec:travel},
we discuss the magnetic field and electron distribution along the line-of-sight.  In Section \ref{sec:data} the
analysis of the Fermi-LAT data of PKS $2155-304$ to constrain the ALP
parameters is given. The results are presented in Section \ref{sec:result}.
Our conclusions are summarized in Section \ref{sec:Summary}, together
with some discussions.

\section{ALP-photon oscillation model}\label{sec:model}

The ALP-photon oscillation system can be described by the following
Lagrangian~\cite{alpold}:
\begin{eqnarray}
\mathcal{L}=-\frac{1}{4} F_{\mu\nu}F^{\mu\nu}+\frac{1}{2}\left( \partial_\mu a \partial^\mu a-m_a^2 a^2 \right)-\frac{1}{4} g_{a\gamma} a F_{\mu\nu}\widetilde{F}^{\mu\nu} + \frac{\alpha^2}{90m_e^4}\left[\left( F_{\mu\nu}F^{\mu\nu} \right)^2+\frac{7}{4}\left( F_{\mu\nu}{\widetilde F}^{\mu\nu} \right)^2\right],
\end{eqnarray}
where $g_{a\gamma}$ is the coupling constant between ALPs and photons
which has been found to be smaller than $10^{-10}$ GeV$^{-1}$ for the ALP mass
$m_a\lesssim 1\mu {\rm eV}$. The last term is the
Euler-Heisenberg-Weisskopf (EHW) effective Lagrangian~\cite{EHW} from
the vacuum polarisation which is negligible as long as the magnetic
field is much lower than the critical field strength $B_c=m_e^2/e\approx
4.41 \times 10^{13}$ G.

We describe the photon-ALP beam using the density matrix
$ \rho \left(s\right)$ that is defined as
\begin{eqnarray}
\rho \left(s\right)=
\begin{pmatrix}
A_1(s)
\\
A_2(s)
\\
a(s)
\end{pmatrix}
\otimes
\begin{pmatrix}
 A_1(s)& A_2(s) & a (s)
\end{pmatrix}^*,
\end{eqnarray}
where $s$ is the position along the path of the photon-ALP beam in direction
${\bf e}_3$, and $A_1,~A_2$ are the photon transverse polarizations along axes
${\bf e}_1,~{\bf e}_2$, respectively. In the short-wavelength limit where
$E \gg m_a$, we  get the Liouville-Von Neumann (LVN) equation
\begin{eqnarray}
i\frac{d\rho(s)}{ds}=[\rho,\mathcal{M}]
\end{eqnarray}
with the mixing matrix $\mathcal{M}$
\begin{eqnarray}
\mathcal{M}_{{\bf B_T}\parallel {\bf e}_2}=\begin{pmatrix}
 \Delta_{pl } &  0&0 \\
 0& \Delta_{pl }  & \Delta_{a \gamma} \\
 0&\Delta_{a\gamma}  & \Delta_{aa}
\end{pmatrix},
\end{eqnarray}
where ${\bf B}_T$ is the transversal magnetic field orientated along
${\bf e}_2$, and
\begin{eqnarray}
\Delta_{a\gamma}&=&\frac{1}{2}g_{a\gamma}B_T\simeq 1.52 \times 10^{-2} g_{a\gamma,-11} B_{T,-6}\,{\rm kpc}^{-1},\\
\Delta_{aa}&=&-\frac{m_a^2}{2E}\simeq -7.8 \times 10^{-2} m_{a,-9}^2 E_{9}^{-1}\,{\rm kpc}^{-1},\\
\Delta_{pl}&=&-\frac{w_{pl}^2}{2E}\simeq -1.1 \times 10^{-4}E_{9}^{-1} \left( \frac{n_e}{\rm cm^{-3}}  \right)\,{\rm kpc}^{-1},
\end{eqnarray}
where $w_{pl}$ is plasma frequency and $n_e$ is the number density of free electrons,
$B_T=B_{T,-6}10^{-6}$ G, $g_{a\gamma}=g_{a\gamma,-11}10^{-11}\,
{\rm GeV}^{-1}$, $m_a=m_{a,-9}\,{\rm neV}$, and $E=E_{9}\,{\rm GeV}$.

For a general magnetic field with angle $\psi$ from ${\bf e}_2$, the mixing
matrix becomes
\begin{eqnarray}
\mathcal{M}=V\mathcal{M}_{{\bf B_T}\parallel {\bf e}_2}V^T,
\end{eqnarray}
with
\begin{eqnarray}
V=\begin{pmatrix}
\cos\psi  & \sin\psi & 0\\
-\sin\psi  & \cos\psi  & 0\\
 0 &  0 &1
\end{pmatrix}.
\end{eqnarray}

Finally, for an initial photon matrix $\rho(s_0)$ at $s_0$, we can get the
density matrix at position $s$ by splitting the path to small regions where
the magnetic field is approximately constant and homogeneous, as
\begin{eqnarray}
\rho(s)=T(s,s_0)\rho(s_0)T^\dagger(s,s_0),
\end{eqnarray}
where $T(s,s_0)$ is the transfer function defined as $T(s,s_0)=\prod^{n}_{0}
T(s_{i+1},s_i)$.

It is helpful to consider a simple case where the constant and homogeneous
magnetic field satisfies ${\bf B}_T \parallel {\bf e}_2$. In this case,
the conversion probability of an initially ${\bf e}_2$ polarized photon
to an ALP is
\begin{eqnarray}
P_{a\gamma}(s,s_0)=\sin^2(2\theta)\sin^2\left(\frac{\Delta_{\rm osc}\left(
s-s_0 \right)}{2}\right),
\end{eqnarray}
where the mixing angle $\theta$ and oscillation wave number
$\Delta_{\rm osc}$ are
\begin{eqnarray}
\theta &=& \frac{1}{2} \arctan \left(\frac{2\Delta_{a\gamma}}{\Delta_{pl}-\Delta_{aa}} \right),\\
\Delta_{osc} &=& \left[\left(\Delta_{pl}-\Delta_{aa} \right)^2+4\Delta_{a \gamma}^2\right]^{1/2}.
\end{eqnarray}
In order to get a significant ALP-photon oscillation, the phases of the
two sine functions should be large enough. 
Firstly the length of the propagation path $l$ should be larger than the
ALP-photon oscillation wave length $l_{\rm osc}=4\pi/\Delta_{\rm osc}$.
In a simple case where $\Delta_{pl}$ and $\Delta_{aa}$ are smaller than
$\Delta_{a\gamma}$, we get $l_{\rm osc}=2\pi/\Delta_{a\gamma}$. Secondly,
the energy should be higher than the lower critical energy $E_L$ which reads
\begin{eqnarray}
E_L &=& \frac{E(\Delta_{aa}-\Delta_{pl})}{2\Delta_{a\gamma}}\simeq \frac{2.5 | m^2-\omega_{pl}^2  |}{\left(\,{\rm neV}\right)^2}B_{T,-6}^{-1}g_{a\gamma,-11}^{-1}\,{\rm GeV}.
\end{eqnarray}

\section{PHOTON TRAVEL ENVIRONMENTS }\label{sec:travel}
To get the survival probability of a $\gamma$-ray photon which travels
from the source to the detector, we need to consider its propagation in
four main regions: the source region (surrounding PKS $2155-304$),
the inter-cluster region, the extragalactic space, and the interior of
the Milky Way.

In the source region, the magnetic field and electron density are
$B\approx 70 \left(r/{\rm pc} \right)^{-1.31}~\mu$G and
$n_e\approx 2 \times 10^4 \left(r/{\rm pc} \right)^{-1.35}$ cm$^{-3}$,
respectively \cite{agnmag}. With the ALP parameter range considered in this
work (i.e., $10^{-12}~{\rm GeV^{-1}}<g_{a\gamma} <10^{-10}~{\rm GeV^{-1}}$,
$10^{-1}~{\rm neV}<m_a<10^2~{\rm neV}$), the oscillation in the source
region is negligible.

There are no observations about the inter-cluster magnetic fields (ICMFs)
around PKS $2155-304$. The typical magnetic field strength in galaxy
clusters is in the range of $1-10~\mu$G \cite{icm}. Following Ref.
\cite{random}, we describe the ICMFs as Gaussian turbulent fields with
mean value zero and variance $\sigma_B$. The power spectrum follows a
power-law $M(k)\propto k^{q}$ ($k$ is the wave number) with $k_{L}<k<k_{H}$
($k_{L}=2\pi / \Lambda_{\rm max}$, $k_{H}=2 \pi / \Lambda_{\rm min}$).
The radical profile of the magnetic field strength is
$B_{r}=B_0[n_e(r)/n_e(0)]^\eta$. We set the maximum radii as
$r_{\rm max}=350$ kpc, beyond which the ICMF is zero. The modified King
model for the electron density distribution is assumed as in Ref. \cite{icm,icm-e},
i.e., $n_e(r)=n_0(1+r^2/r_c^2)^{-3\beta/2}$, with parameters $n_0=0.01$
cm$^{-3}$, $r_c=200$ kpc, $\beta=2/3$. Other parameters are given in
Table \ref{tab1}.

\begin{table*}[]
	\caption{Parameters of the ICMF.}
	\begin{tabular}{ccccc}
		\hline
		$\sigma_B$ & $\Lambda_{\rm min}$ & $\Lambda_{\rm max}$ & $q$ & $\eta$\\
		\hline
		$3.0~\mu$G & 0.5 kpc & 20 kpc & $-11/3$ & 0.5\\
		\hline
	\end{tabular}
	
	\label{tab1}
\end{table*}

 The extragalactic magnetic fields (EGMFs) are not clear either.
	The upper limit on the EGMFs given in literature is about several nG and the real value is expected to be much lower
	\cite{ebl1,ebl2,ebl3,ebl4,ebl5}. With $B_T=1~{\rm nG}$, the lower criterion is about $E_L=2.5 m_{-9}^2g_{a\gamma,-11}^{-1}\,{\rm TeV}$. The most energetic $\gamma-$rays of PKS 2155-304 recorded by Fermi-LAT are well below $\sim 1~{\rm TeV}$. Therefore, we ignore the photon-ALP mixing
	in the EGMFs.  The $\gamma$-rays would also be absorbed by the extragalactic
	background light (EBL). In this work we adopt the EBL model presented in Ref.~\cite{ebl}.

  For the magnetic
	fields in the Milky Way, we adopt the model developed in Ref.~\cite{mw-GMF}.
	The Milky Way magnetic fields have both regular and striated random components. As the coherent length of the random
	component (100 pc or less) \cite{GMF-len} is too short to have a significant
	contribution, we only take into account the regular component.  The Milky Way electron density model developed in Ref.~\cite{mw-ele}
	is adopted.

With the information of magnetic fields, we can get the final density matrix
transported from an initial pure photon beam $\rho(s_0)=1/2 ~ {\rm diag}
(1, 1, 0)$ as
\begin{eqnarray}
\rho(s)=T_{\rm MW}T_{\rm EBL}T_{\rm ICMF}(s,s_0)\rho(s_0)\left(T_{\rm MW}
T_{\rm EBL}T_{\rm ICMF}\right)^\dagger(s,s_0),
\end{eqnarray}
where $T_{\rm MW}$ and $T_{\rm ICMF}$ are the transfer functions in the
Milky Way and the inter-cluster magnetic fields, respectively, and $T_{\rm EBL}={\rm diag}
(e^{-\tau/2}, e^{-\tau/2}, 1)$ takes the EBL attenuation effect into account
($\tau$ denotes the optical depth of $\gamma$-rays). Then the photon
survival probability is given by
$P_{\gamma \gamma}=\rho(s)_{1,1}+\rho(s)_{2,2}$.

\section{Fermi-LAT observation of PKS $2155-304$}\label{sec:data}

Our data analysis procedure is similar to that in
Ref.~\cite{Fermi}. Considering the high statistics of emission from
PKS $2155-304$, we use only the EDISP3\footnote{See \url{https://fermi.gsfc.nasa.gov/ssc/data/analysis/documentation/Cicerone/Cicerone_Data/LAT_DP.html}
for the description of the event types in the Fermi-LAT data.} data
to perform the analysis to achieve the best energy resolution.
We follow the standard thread recommended by Fermi Collaboration\footnote{\url{https://fermi.gsfc.nasa.gov/ssc/data/analysis/documentation/Cicerone/Cicerone_Data_Exploration/Data_preparation.html}} to filter the data.
The SOURCE data from 2008-10-27 to 2017-06-12 (MET 246823875-518983875)
for energies between 100 MeV and 500 GeV with zenith angle
$\theta_{\rm z}<90^\circ$ are selected. The selected data are binned
into a count cube with $100\times100$ spatial bins (0.1$^\circ$ bin size)
and 145 energy bins\footnote{The widths of energy bins are roughly 30\% of
the median energy resolution of the EDISP3 data, see Ref.~\cite{Fermi}
for details.} to perform a binned likelihood analysis. We fitted the
model to the data over the entire energy range ({\tt global fit}) to
determine the parameters of background sources. In this step, the
spectral parameters of all point sources (including the 3FGL sources
and new point sources with TS$>25$ in the residual TS map) within a
$5^\circ$ radius are left to be free. We also left the normalizations
of the two diffuse background emission components, {\tt gll\_iem\_v06.fits}
and {\tt iso\_P8R2\_SOURCE\_V6\_v06.txt}, free. The background parameters
were then fixed to their best-fit values in the subsequent likelihood
profile analysis.

The likelihood profile in each energy bin, ${\mathcal L}_{k'}(\mu_{k'}
(P_{k'})|D_{k'})$, is obtained by varying the prefactor $P_{k'}$ of the
target source PKS $2155-304$ and calculating the corresponding likelihood
${\mathcal L}$. Here $D_{k'}$ denotes the observed data in the $k'$-th
bin, and $\mu_{k'}$ is the expected number of photon counts which is a
function of $P_{k'}$. As mentioned above, all background parameters are
fixed in the procedure. The spectral index of PKS $2155-304$ is fixed
to be 2.0. We have tested that for such small energy bins, varying the
index from 2.0 to 2.5 does not affect the results.

Given the photon survival probability for one ICMF realization ${\rm B}_i$,
the expected number of photon counts is
\begin{eqnarray}
\mu_{k'}=\sum_{k} D_{k k'} \int_{\Delta E_{k}} dE P_{\gamma \gamma}
F(E) \varepsilon(E),
\end{eqnarray}
where $D_{k k'}$ is the energy dispersion which converts the
theoretical numbers of photon counts in the true energy bins $\Delta E_{k}$
to the expected numbers of photon counts in the reconstructed energy bins
$\Delta E_{k'}$, $\varepsilon(E)$ is exposure for true energy, $P_{\gamma\gamma}$ is the photon survival probability and $F(E)$ is the spectrum of PKS $2155-304$.

We considered three different types of spectral models for PKS $2155-304$,
including {\tt Power Law}, {\tt Broken Power Law}, and {\tt Log-Parabola}.
The log-likelihood values of the global fit for these three models are
${\rm log}({\mathcal L})=-59269.9$, $-59261.4$, and $-59260.7$,
respectively.
Thus, we used the {\tt Broken Power Law} model of PKS $2155-304$ hereafter,
which reads
\begin{eqnarray}
F(E)=\left\{\begin{matrix}
 & N E^{-k_1},     & E<E_{\rm br}  \\
  & N E^{-k_2} E_{br}^{k_2-k_1}, & E \geq E_{\rm br}
\end{matrix}\right.,
\end{eqnarray}
where $E_{\rm br}$ is the break energy which is $24$ GeV as obtained from
the global fit.

\begin{figure}[htbp]
\includegraphics[width=1.0\textwidth]{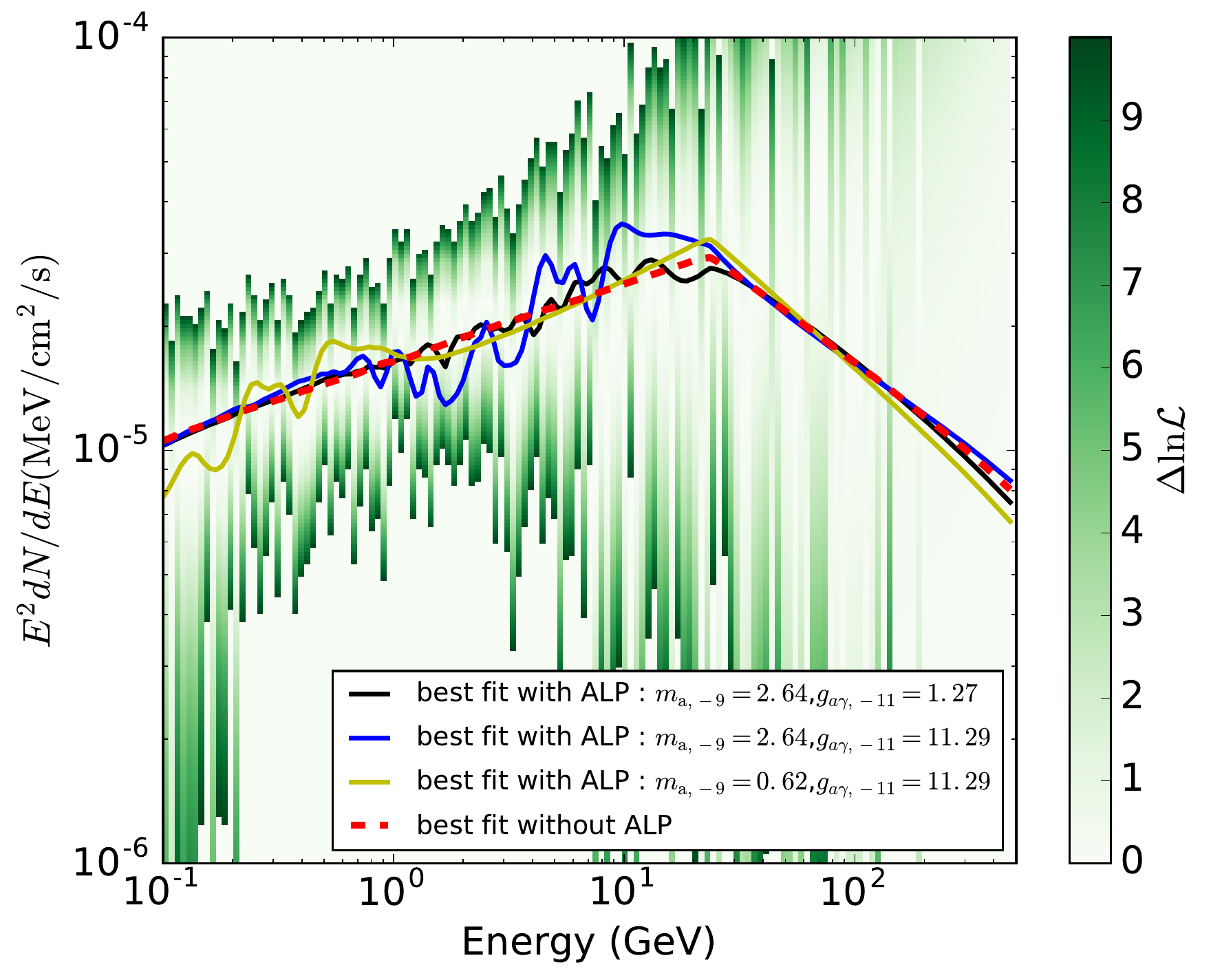}
\caption{Shown in color are the likelihoods, $\Delta \ln \mathcal{L}=
\ln\mathcal{L}_{\rm max}-\ln\mathcal{L}$, on the plane of energy and
flux. The red dashed line shows the best-fit spectrum without ALP,
and the solid lines show the best-fit spectra for three sets of ALP
parameters.}
\label{data}
\end{figure}

For each ALP parameter and realization of the ICMF, we maximized the
joint likelihood of all reconstructed energy bins $\Delta E_{k'}$
\begin{eqnarray}
\mathcal{L}(m_a,g_{a\gamma},B_i,\theta|D)=\prod_{k'}\mathcal{L}
(m_a,g_{a\gamma},B_i,\theta|D_{k'}),
\end{eqnarray}
where $\theta$ stands for the nuisance parameters of the intrinsic
spectrum. In Fig.~\ref{data}, we showed the likelihood ($\Delta\ln
{\mathcal L}$) map on the energy and flux plane, together with the
best-fit model spectra without and with ALP (for three sets of ALP
parameters).

\section{Constraints on ALP parameters}\label{sec:result}

It is challenging to get a well-defined likelihood of an ALP model due to
the random realizations of the ICMFs. We used the Bayesian method to take
the ICMF distribution into account. Flat priors of the ICMF realizations,
$\ln(m_a)$ and $\ln(g_{a\gamma})$ are assumed. In total, 800 realizations
of the ICMFs are generated, and the prior ALP parameter ranges are
$0.1<m_a/{\rm neV}<10^2$ and $10^{-12}<g_{a\gamma}/{\rm GeV^{-1}}<10^{-10}$.
\begin{figure}[htbp]
	\includegraphics[width=0.8\textwidth]{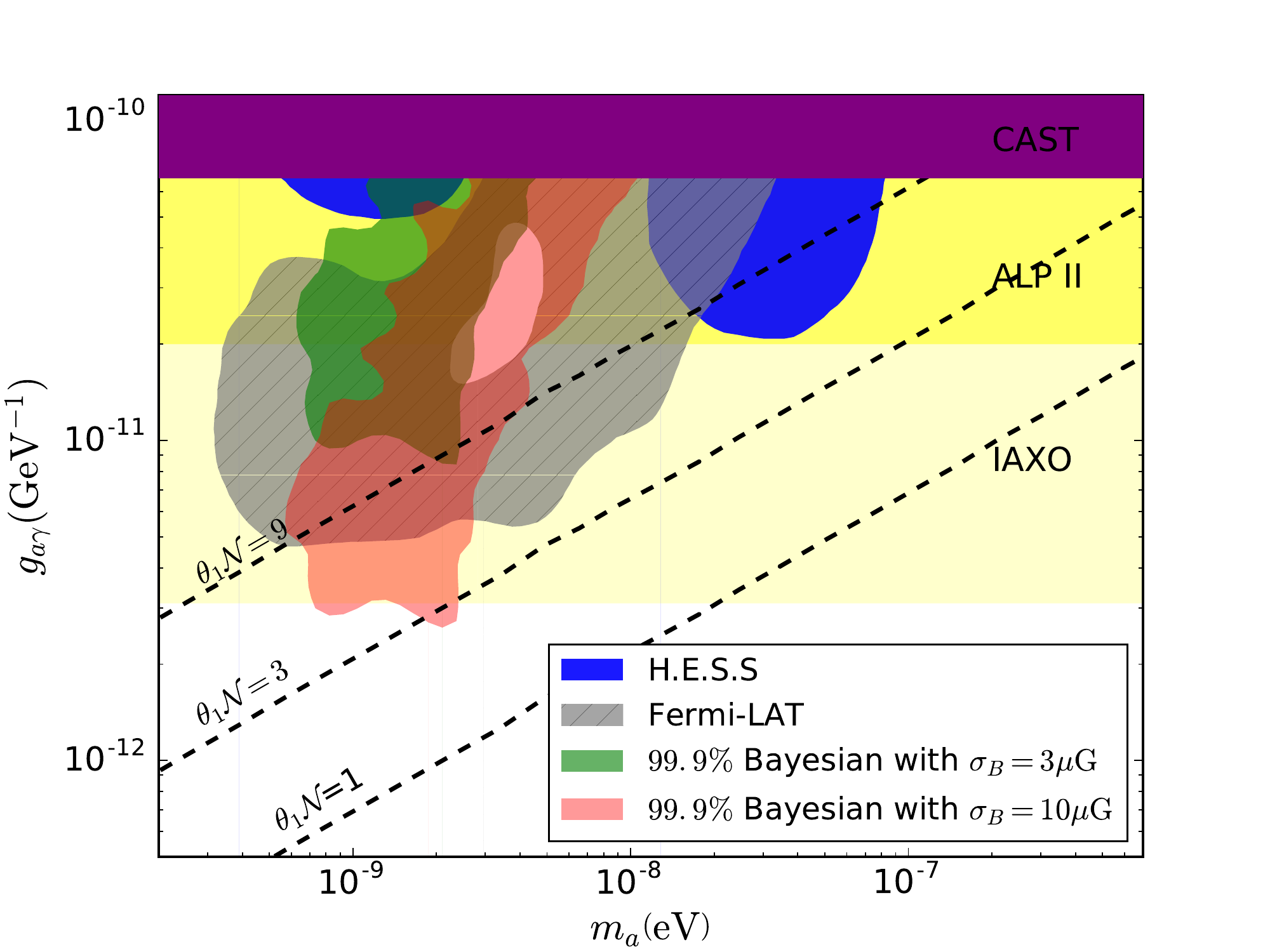}
	\caption{Exclusion regions (red for 99\% and green for 99.9\% credible
		levels) of the ALP parameter space ($m_a,g_{a\gamma}$) derived in this
		work, compared with that obtained by Fermi-LAT observation of NGC
		1275 (gray; \cite{Fermi}) and H.E.S.S. observation of PKS $2155-304$
		(blue; \cite{hess-alp}). The dashed line with $\theta \mathcal{N}=1$
		($\theta_1 \mathcal{N}=3$,$\theta_1 \mathcal{N}=9$) denotes the parameter space that the ALP
		dark matter accounts for all ($1/9$,$1/81$) of the dark matter. Some other
		constrains and sensitivities are adopted from~\cite{limitsplot,
			limitsplot1, CAST2, ALPII,CAST,IAXO,axion3}.
	}
	\label{limit}
\end{figure}

\begin{figure}[htbp]
	\centering
	\subfigure{
		\includegraphics[width=0.20\textheight]{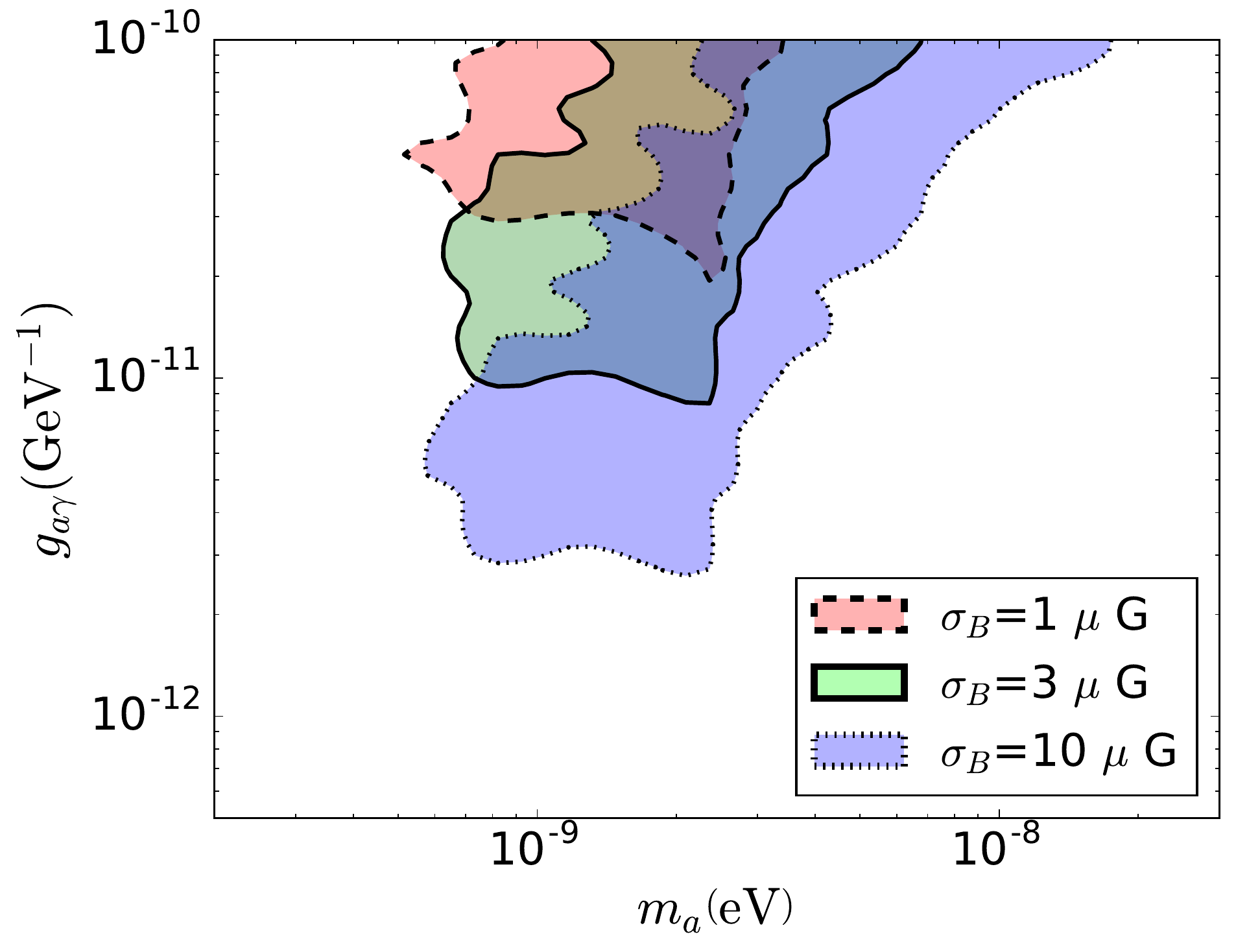}}
	\subfigure{
		\includegraphics[width=0.20\textheight]{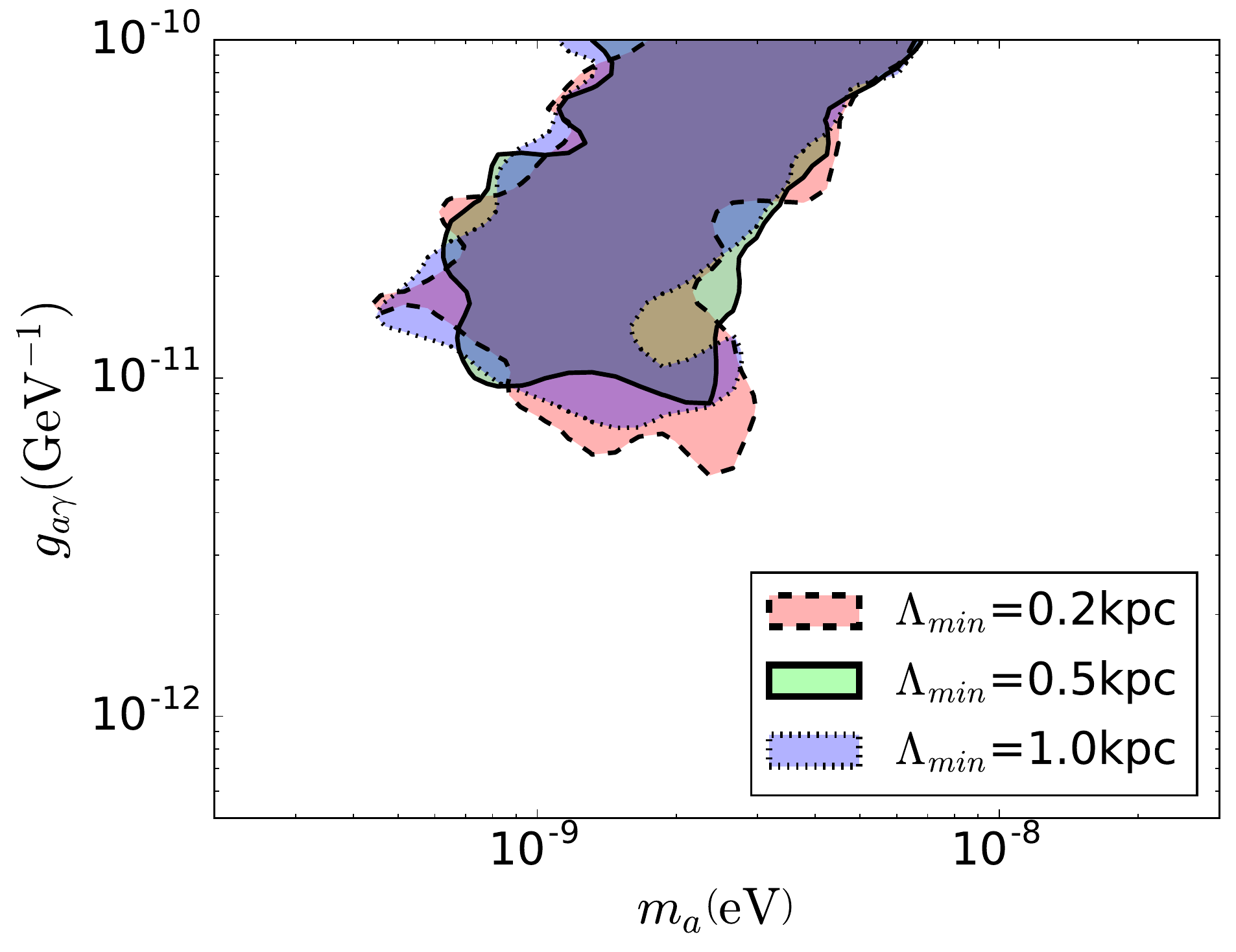}}
	\subfigure{
		\includegraphics[width=0.20\textheight]{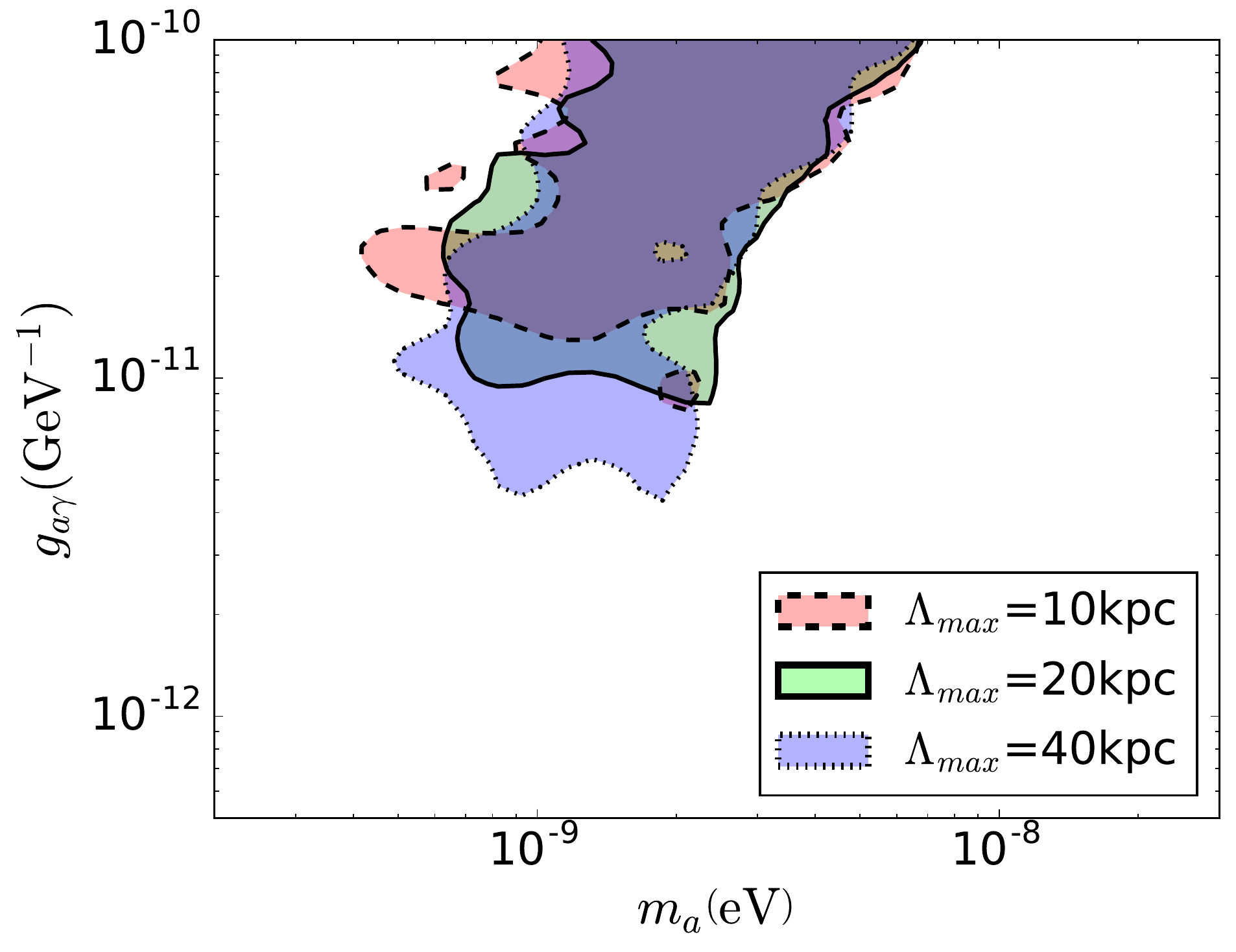}}
	\subfigure{
		\includegraphics[width=0.20\textheight]{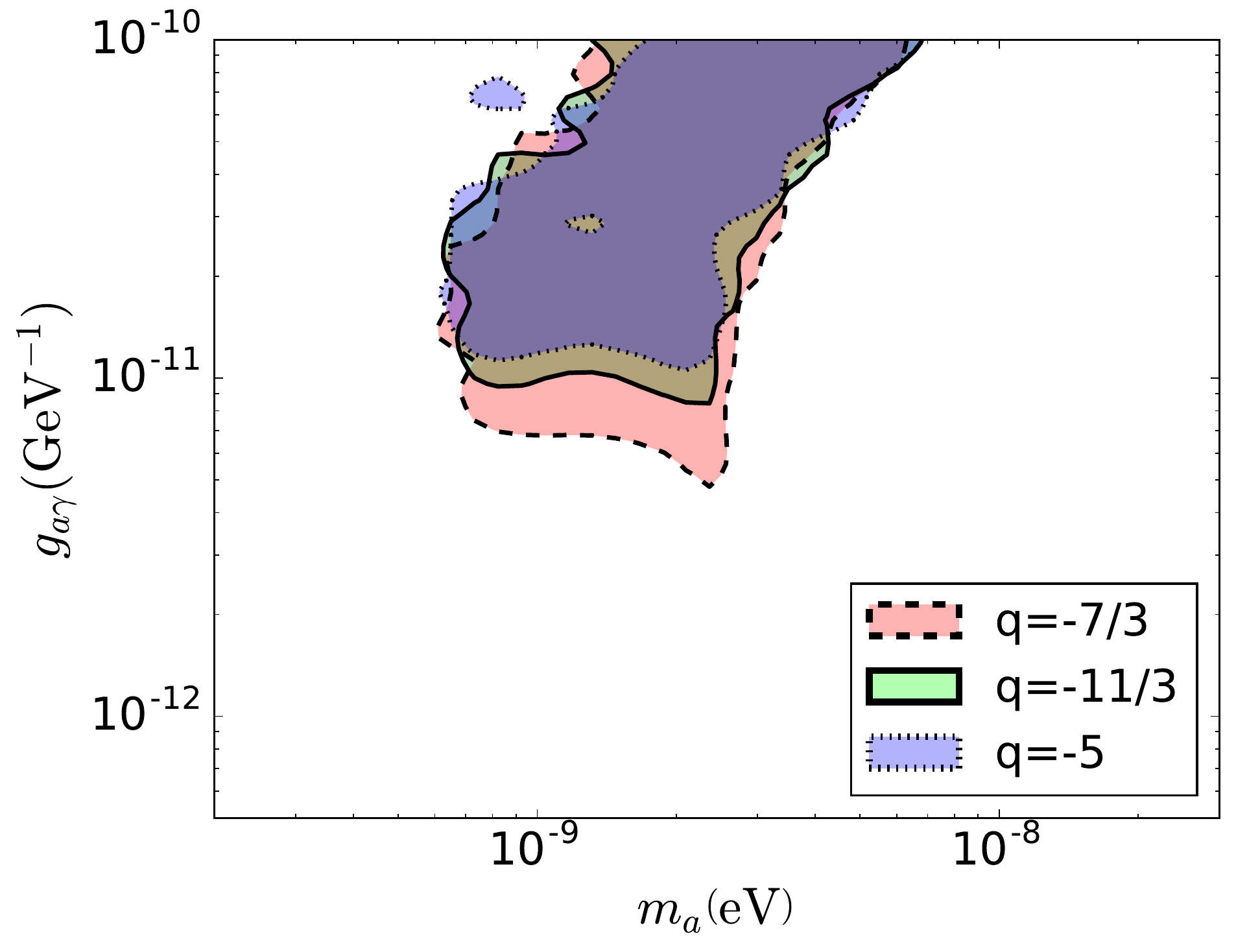}}
	\subfigure{
		\includegraphics[width=0.20\textheight]{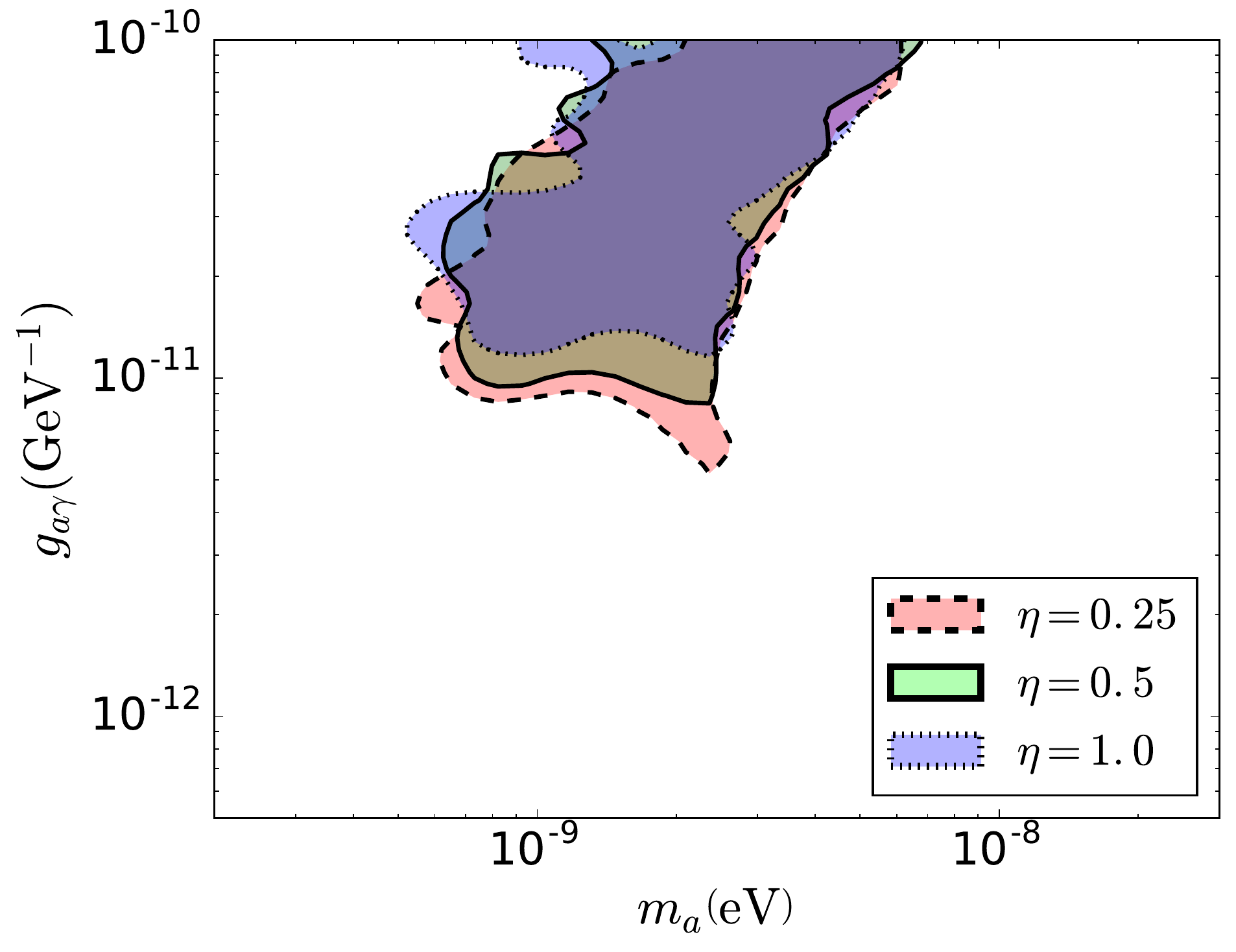}}
	\subfigure{
		\includegraphics[width=0.20\textheight]{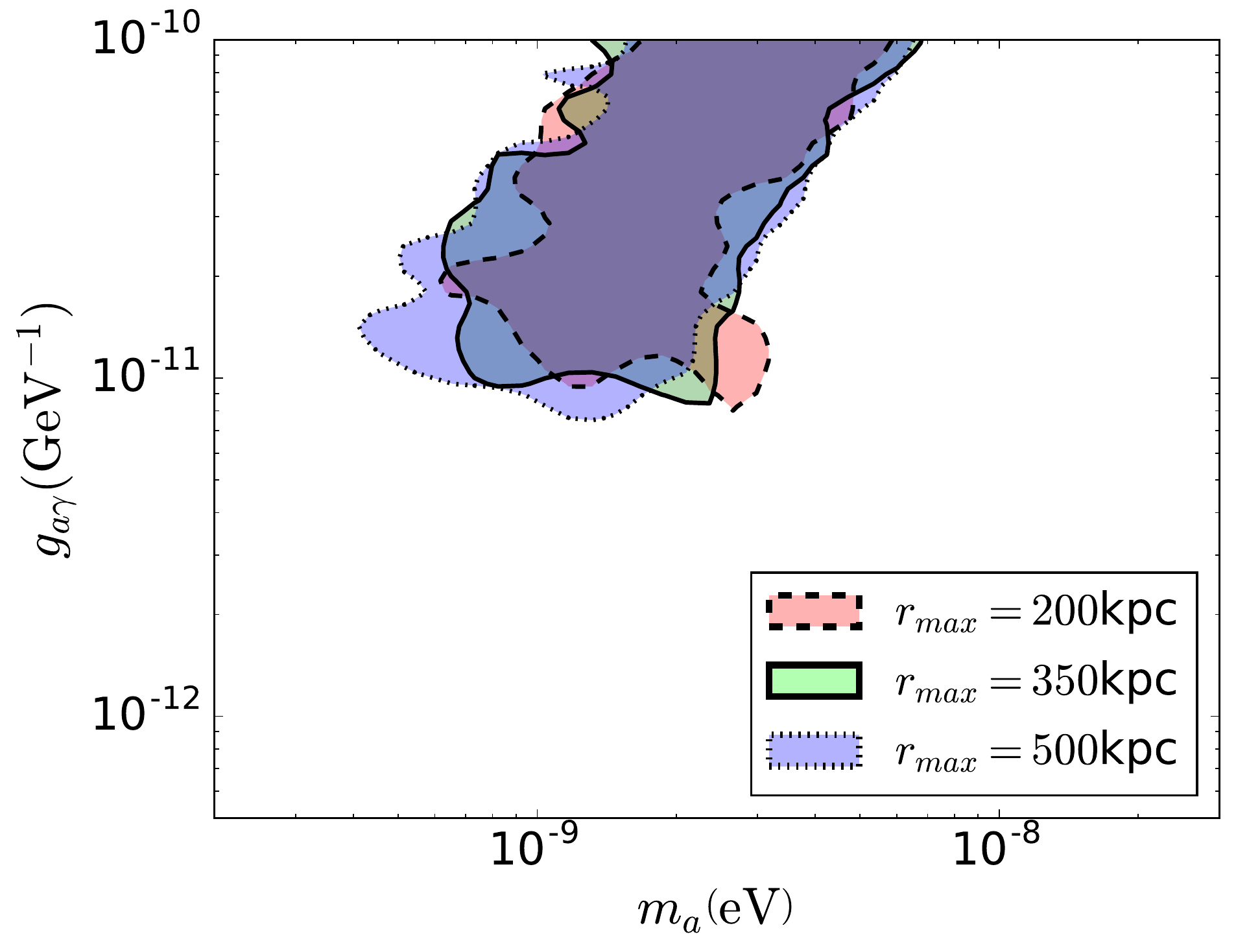}}
	\caption{The $99.9\%$ credible level exclusion parameter regions with different ICMF parameter values. The green region with solid line is the one with the fiducial parameters given in Table \ref{tab1}.}\label{scan}
\end{figure}

 The $99.9\%$ credible level exclusion parameter regions obtained
in this work are shown in Fig.~\ref{limit}. For comparison, we also presented the
limits from previous works.  With the parameters given in Table \ref{tab1}, the exclusion region (the green one in Fig.~\ref{limit}.) is a strip region of
${\rm 0.6~neV < m_a < 4~neV}$ and $g_{a\gamma} > 10^{-11}~{\rm GeV}^{-1}$.
For $\sigma_B=10~\mu {\rm G}$, the tightest constraint on $g_{a\gamma}$
reaches about $3 \times 10^{-12}~{\rm GeV}^{-1}$.
In such a case, our exclusion region
is comparable to (though a bit narrower than) that set by the Fermi-LAT
collaboration~\cite{Fermi} and covers the ``hole"-like area.
For $\sigma_B \sim 1-3~\mu {\rm G}$, the constraints are weaker
(see Fig. \ref{scan}). Such behaviors are actually anticipated
(see Sec. \ref{sec:model}). As shown in Fig. \ref{scan}, the exclusion
regions also depend on $\Lambda_{\rm max}$. The other parameters do not
affect the results significantly.

Future experiments such as ALP II~\cite{ALPII} and IAXO~\cite{IAXO} will
be able to probe ALPs in a wider parameter space. As shown in Fig.~\ref{limit},
some reachable regions by such experiments have already been excluded
by the current $\gamma$-ray observations. The dashed lines shown in
Fig.~\ref{limit} labeled as $\theta_1 \mathcal{N}=1$, $\theta_1 \mathcal{N}=3$,
$\theta_1 \mathcal{N}=9$ are the parameters for the ALP dark matter
making up $100\%$, $1/9$, $1/81$ of the dark matter in the
Universe~\cite{ALP-dm}. The constraints derived in this work can
exclude some of the parameter region of the ALP dark matter.


\section{Summary}
\label{sec:Summary}

The ALP-photon oscillation is caused by the direct interaction between
the ALP and electromagnetic fields, which could induce spectral irregularities
of high energy $\gamma$-ray sources due to the presence of magnetic
field in the Universe. In this work, we search for such spectral
irregularities in the $\gamma$-ray spectrum of PKS $2155-304$ using 8.6
year Fermi-LAT observations. We do not find significant evidence for ALPs.
Constraints on the parameters of ALPs are thus obtained.

We have ruled out  $g_{a\gamma} > 10^{-11}~{\rm GeV}^{-1}$ for ALP mass
of $0.6~{\rm neV} <m_a<4~{\rm neV}$ at the 99.9\% credible level.
If the ICMFs around PKS $2155-304$ are as strong as $10~\mu {\rm G}$, the
``hole"-like region allowed in previous work of Ref.~\cite{Fermi} is excluded.
As demonstrated here and in the literature, the high energy resolution
observations of $\gamma$-ray spectra can effectively probe ALPs.
The forthcoming high energy resolution data of
the Dark Matter Particle Explorer~\cite{DAMPE} and the future High Energy
cosmic-Radiation Detection facility~\cite{HERD,Huang:2015fca} are expected
to significantly advance such studies.

\acknowledgments
This work was supported in part by the National Key Research and Development
Program of China (Nos. 2016YFA0400200 and 2016YFE0129300), the National
Natural Science Foundation of China (Nos. 11535004, 11120101005, 11175085,
11773075), the Youth Innovation Promotion Association of Chinese Academy
of Sciences (No. 2016288), the Natural Science Foundation of Jiangsu Province
(No. BK20151608), and the Science and Technology Development Fund of Macau
(No. 068/2011/A).

\end{document}